\documentclass[11pt]{amsart}
\usepackage{geometry}                
\usepackage{graphicx}
\usepackage{amssymb}
\usepackage{epstopdf}
\title{Signal Propagation In Double Edged Relays }
\author{Adam Boucher}
\date{Feburary 23, 2023}                                           

\begin{document}
\begin{abstract}
A discrete signal propagation model blending characteristics of linear wave propagation and finite state automata is developed. We show this model obeys a limited form of superposition and is capable of displaying a wide variety of interesting behaviors. We show how the model's superposition properties permit information to be encoded and retained by signals that pass through discrete networks. Applications to combinatoric problems are outlined.  \end{abstract}

\maketitle

\section{Introduction} 

This paper defines and develops the theory of signal propagation in double edged relays. We will refer to various configurations of double edged relays as SPIDER models. This acronym is derived from the biological inspiration for this work. If you imagine a spider waiting patiently on its web, it can sense vibrations along the strands of the web and use those vibrations to obtain information about the type and location of prey and other disturbances anywhere in its web. As any frustrated childhood arachnologist can tell you spiders appear to be able to distinguish between different kinds of incident vibrations within the web and are not often fooled by curious children trying to trick the spider into exploring the web.\\
A double edged relay mixes the formalism of finite state automata with the discretized dynamics of the one dimensional wave equation. By blending these two mathematical frameworks we are able to create a very flexible framework for creating signaling simulations which can reveal many interesting features of graphs and networks. By creating double edged relays which are forced to behave like the linear wave equation we are able to exploit several important well known properties of solutions to linear hyperbolic partial differential equations: finite propagation speed, and linear superposition of signals.\\
The paper begins by reviewing the a limited form of the D'Alembart solution to the wave equation in one spatial dimension. This theory is used to demonstrate how information can be encoded into traveling waves. This introduction is followed by the definition of a double edged relay which we prove obeys a limited form of superposition. (Namely: superposition of signals in different positions on the relay, and superposition of amplitudes.). With these fundamentals in place we demonstrate how using signal propagation in individual arrays can be used to achieve simple computational tasks.\\
With the theoretical foundations in place we explain how to overlay double edged relays over existed graphs and several techniques for encoding information into the amplitude of the signals. These ideas are applied to three well-known graph theoretic problems: In this paper we explore how SPIDER models can solve the shortest path problem of graph theory, and offer a narratively compelling algorithmic alternative to Dijkstra's algorithm.  \\

\section{Review of D'Alembart's formula for a stationary initial displacement.}

The wave equation of mathematical physics is a hyperbolic partial differential equation which has been studied for hundreds of years. As a hyperbolic equation it possesses many amazing properties including linear superposition and finite propagation speed of information. The D'Alembart solution to the wave equation in one dimension has a particularly beautiful form when an elastic string is subjected to a displacement with no initial velocity.\cite{1} To those unfamiliar with the equation or this particular solution:\\

The initial value problem to find $u(x,t)$ in $(x,t)\in (-\infty, \infty)\times (0,\infty)$ given $f(x)$ . 
\begin{align}
	\frac{\partial^2 u}{\partial t^2} -c^2\frac{\partial^2 u}{\partial x^2} = 0, \hspace{.2in} u(0)= f(x), \hspace{.2in} \frac{\partial u}{\partial t}|_{(x,t=0)} = 0, \nonumber\end{align}
has the exact solution:
\begin{align}
	u(x,t) = \frac{1}{2}f(x-ct) + \frac{1}{2} f(x+ct) \nonumber
\end{align}

In the special case where $f$ has compact support (i.e. is identically zero outside a finite interval) this eventually resolves into two disjoint waves which appear to have the same shape as the initial displacement, but move in opposite directions at constant speed.\\
\begin{center}
\includegraphics[height=150pt]{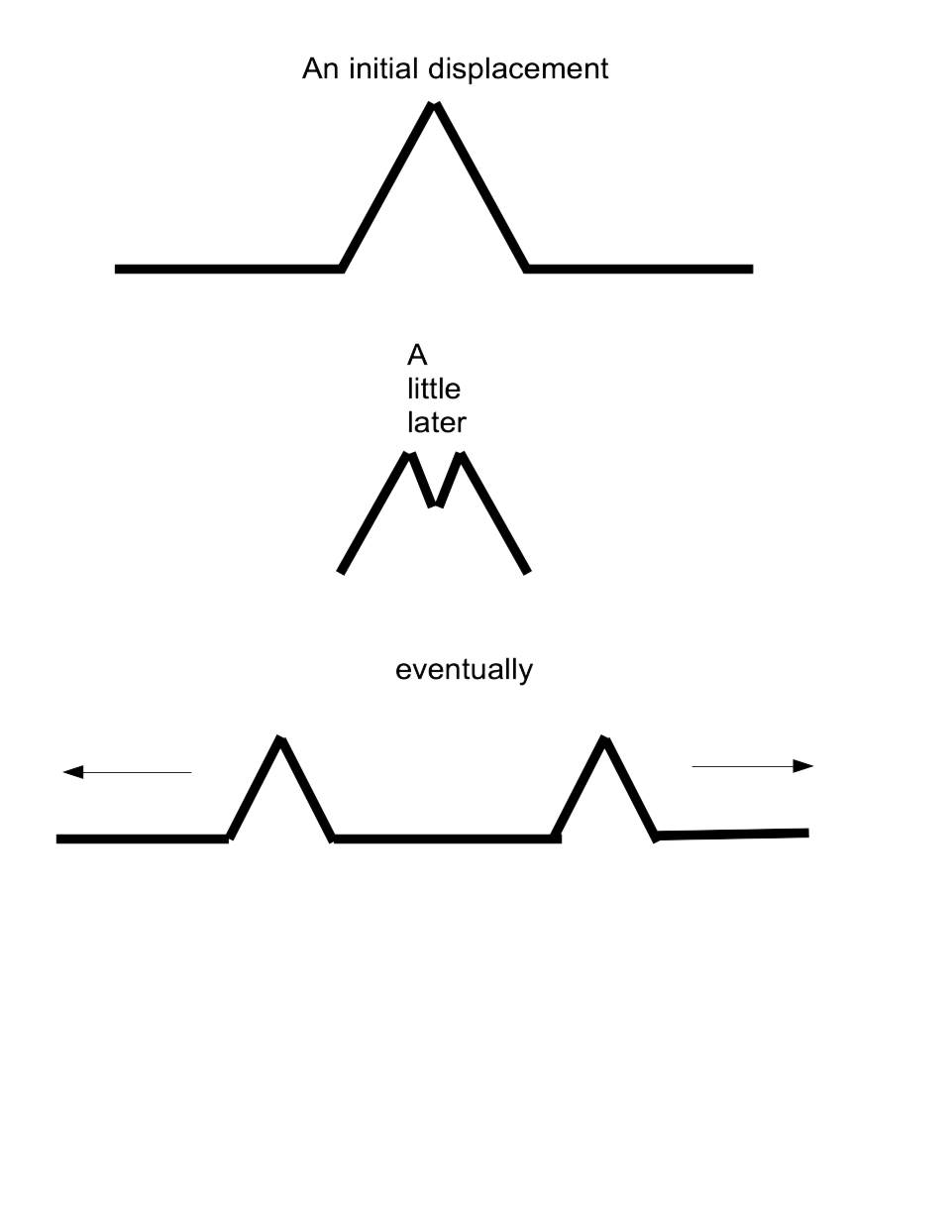}\end{center}
This behavior is robust and can be observed in simulation and experiment (until the solution decays due dissipative forces which are neglected in the theoretical model). Because of this property, one can encode information into any functional shape and transmit it from one location to another along a wave packet. The approach we will adopt here is to pick a single signal shape and encode information into the amplitude of the signal making it stronger or weaker as needed.\\

In 1929 (reprinted in English in 1967) Courant, Friedrichs and Levy developed the theory of the difference equations of mathematical physics, and made the remarkable observation that because of the constant propagation speed of the linear wave equation it is actually possible to create solutions from $\delta$ impulses which are exact solutions to both the continuous and discretized equations. These solution may be simulated with zero error when the grid spacing for the finite difference methods is chosen to match the timestep.\cite{2} For the applications we envision, it is sufficient to limit our attention to unit propagation speed.

\newpage
\section{The SPIDER model}
The double edged relay is created by using paired, equal length discrete arrays to link what we will call 'relays'. In all the examples and applications we develop the relays will be coincident with the vertices of graphs, however the purpose of the relay is to perform operations on the signals carried by the connecting arrays. By changing these operations one may be able to create spider models for myriad applications we have not yet envisioned.\\

One simple instance of a double edged relay is given below:\\

\begin{center}
\includegraphics[height=150pt]{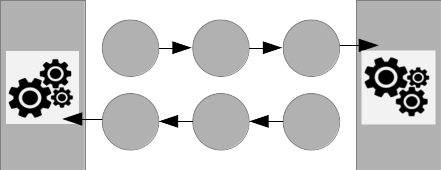}
\end{center}
We restrict the behavior of the edges to follow the transport properties of the discretized wave equation. By limiting ourselves to integer edge weights and delta signals we ensure that the simulations for spider models blend exact solutions to both the continuous wave equation and discretize wave equation. This elegant blending of discrete and continuous creates a compelling model with many exploitable mathematical properties. Each time step we move the array entries exactly one position (according to the arrows). This behavior ensures that along the array signals travel with constant speed, and obey all of the expected superposition properties of the discrete wave equation. By modeling two edges and giving them opposite propagation directions we ensure that the relays can communicate with each other.\\

When a signal reaches a relay we are free to specify any desired operation including: transport, amplifying, splitting, copying, filtering. A relay could have a single or multiple functions and could change its behavior based upon any specified criteria. The only restriction for the operation of the relays is that any output signals are computed and transported to the next edge location(s) at the beginning of the time step after the incident signal reached the relay.\\

This restriction ensures that signals traveling through one or more double edged relays maintain the constant propagation speed of the wave equation. This restriction ensures that the time step is always a reliable measure of the distance a signal (possibly mutated) has traveled through the array.\\

To illustrate some of the possible behaviors we have designed double edged relays to achieve two simple programming tasks. We have tried to explain the relay operations graphically.
\begin{enumerate}
\item A periodic signal loop:
\begin{center}
\includegraphics[height=75pt]{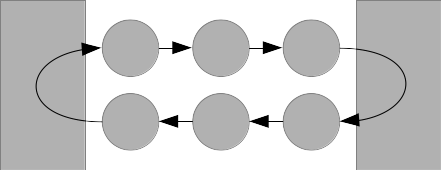}
\end{center}
Here regardless of signal amplitude we simple bounce signals from one edge to the other. This type of relay would precisely mimic a homogeneous Neumann boundary condition.

\item An amplitude alternating loop:
This single relay will create a periodic signal which alternates in sign as it passes from one edge to another. Several time-steps are shown so the reader can follow the logic through the relay.
\begin{center}
\includegraphics[height=75pt]{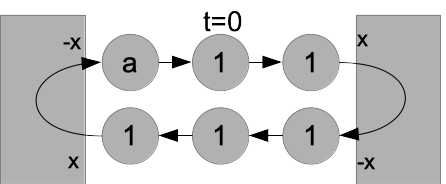}
\includegraphics[height=75pt]{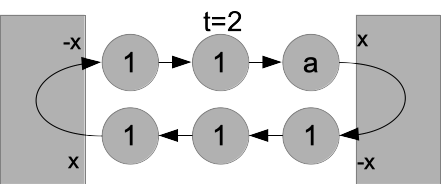}
\includegraphics[height=75pt]{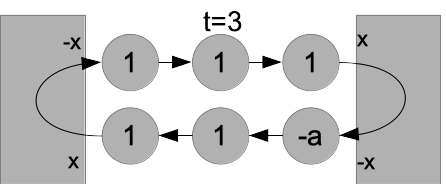}
\includegraphics[height=75pt]{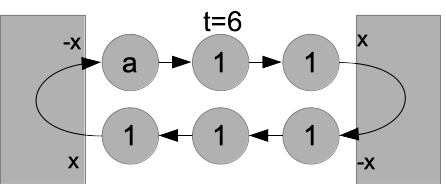}
\end{center}
 This type of condition mimics a traveling wave packet interacting with a homogeneous Dirichlet boundary condition on a finite domain.

\item A signal which degrades after a finite number of steps. Here the operations performed by the relays have a direct impact on the signal amplitudes and result in non-linear behavior.
\begin{center}
\includegraphics[height=75pt]{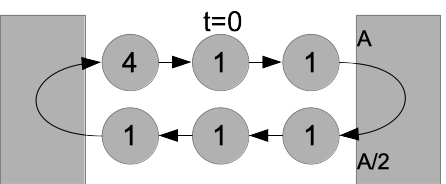}
\includegraphics[height=75pt]{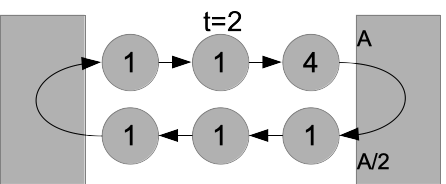}
\includegraphics[height=75pt]{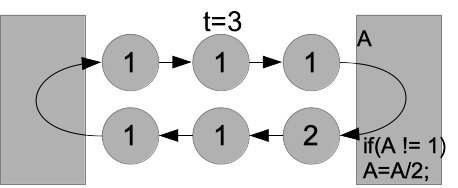}
\includegraphics[height=75pt]{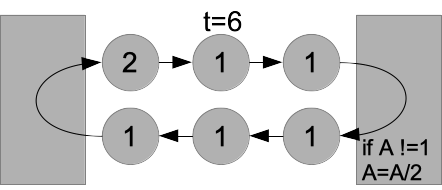}
\end{center}
{\bf{Remark:}} If we amplified the signal amplitude here rather than limit it we would create a signal which evolved according to a geometric progression.\
\end{enumerate}
\section{SPIDERs on graphs}

The most straight forward application of the SPIDER model is to provide a dynamic simulation framework to conduct numerical experiments on graphs. In order to overlay double edged relays over an undirected arbitrary graph (if the graph is weighted we assume the weights are positive integers) we model the vertices of the graph as the relays and when two vertices are connected we model the graph edges with double edge arrays with length equal to the original edges. The figure below shows the configuration for a single vertex drawn with 4 incident edges and empty signals in a discrete radius of 2 away from this vertex.
\begin{center}
\includegraphics[height=100pt]{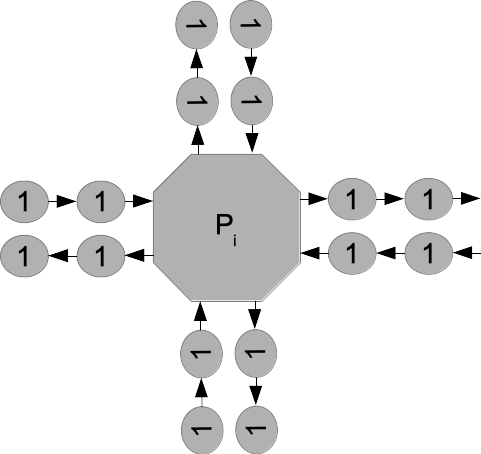}
\end{center}

With this overlay in place we have adopted several conventions which make the underlying mathematics particularly elegant. We model signals traveling through the relays as integers and we treat an amplitude of 1 as an empty signal. We label the vertices of the graph with the prime numbers and we have each vertex use the following relay operations.\\

\begin{enumerate}
\item When a signal with amplitude different from 1 is incident upon an edge we amplify that signal amplitude by the corresponding prime label for that vertex.\\
\item After amplification each vertex transmits the newly amplified signal to all connecting outgoing edges. (Usually it is also useful to filter signals which are revisiting edges they have already visited. This is easy using the prime labeling since we can check newly amplified signals using modulo filtering. We make exceptions to this filtering when completing cycles etc.)
\end{enumerate}
We show the interaction between an incident signal and the vertex in the frames below. In the first illustration we only show amplification and splitting. In the second illustration we show how filtering is employed to stop certain signals from propagating.
\begin{center}
\includegraphics[height=100pt]{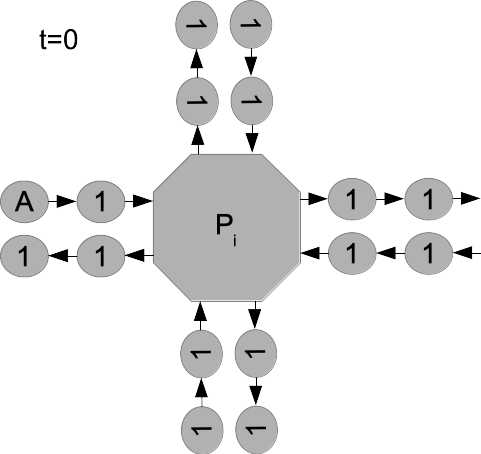}\hspace{.2in}
\includegraphics[height=100pt]{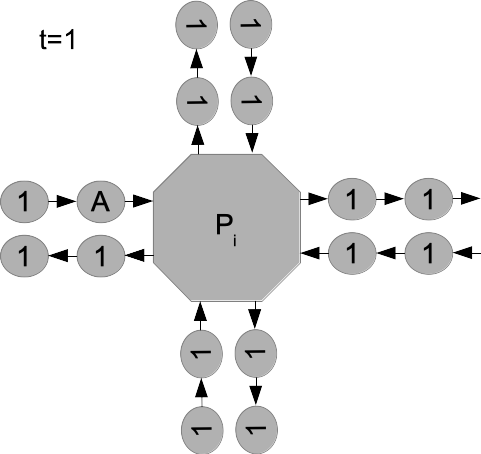}\hspace{.2in}
\includegraphics[height=100pt]{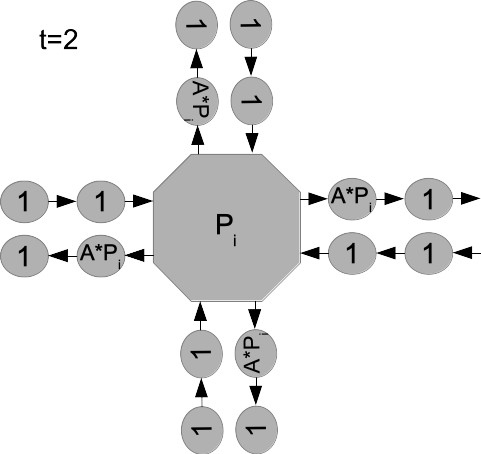} 
\end{center}
Filtering:
\begin{center}
\includegraphics[height=100pt]{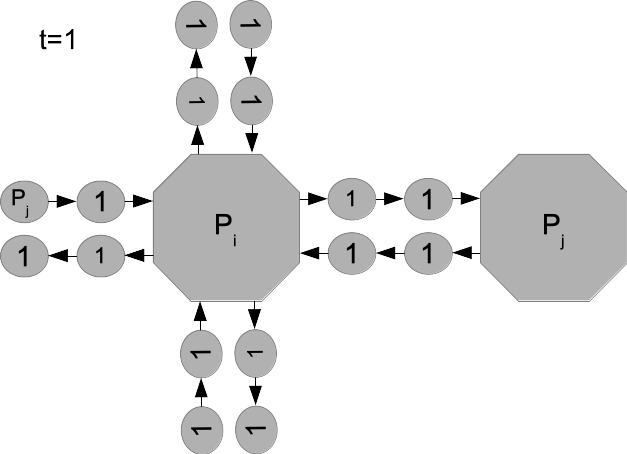}\hspace{.2in}
\includegraphics[height=100pt]{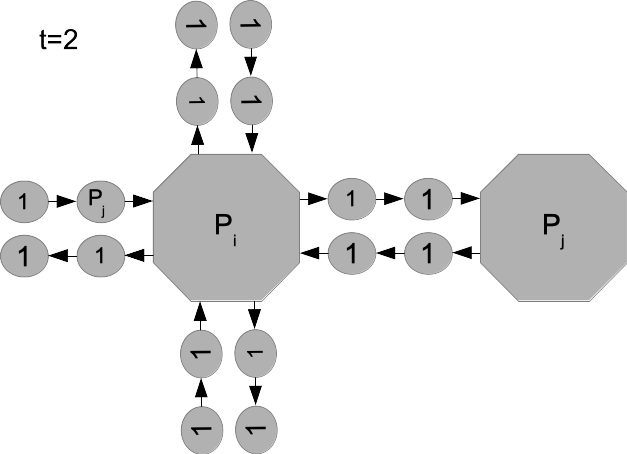}\hspace{.2in}
\includegraphics[height=100pt]{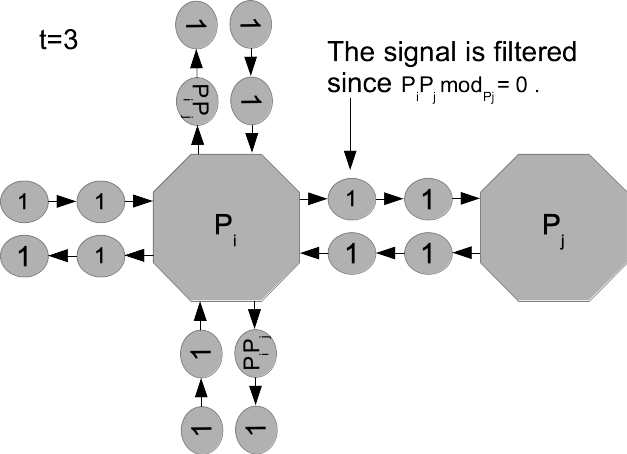} 
\end{center}

This behavior for the edges results in the following emergent behavior: The amplitude of the signals propagating through the graph carry historical information about the vertices visited by that signal. This powerful idea means that SPIDERS can be used to easily generate algorithms for solving combinatorial problems about paths and cycles through graphs.\\

If we consider the shortest path problem for a particular set of vertices in a graph with positive integer edge weights we can create a SPIDER overlay for the graph, and initialize the SPIDER overlay with the prime signal at all edges leaving the starting vertex. We let those signals propagate through the graph until they reach the desired ending vertex. The first signal which reaches the destination vertex corresponds to the shortest path. In the case of any signal collisions (situations where two or more signals are incident upon the same vertex at the same timestep) any of the incident signals can be propagated forward along the edge since all of the incident signals can be traced back to the starting edge using equal length paths.\\
Because of the rigid adherence to the finite propagation speed of the signals the shortest path length is guaranteed to be equal to the number of time-steps of the simulation. By using the prime labeling and multiplicative amplification we can factor the amplitude to obtain a list of all the vertices on the shortest path. (And by analyzing the corresponding signal propagation we can retrieve the ordering of the path as well. This is guarateed since this instatiation of the algorithm splits signals in forward time. Backtracking through the signal history is equivalent to backtracking through a tree.) The factoring here is not difficult since amplitudes are limited to binary combinations of the corresponding vertices. If one wishes to use this model for combinatorial optimization rather than just finding the shortest path between two vertices, then using bit strings recording vertex visitation and pointers to the relay edge locations would provide better scaling for parallel implementation. \\

\section{Discussion and Further Work}

The SPIDER model provides an interesting hybrid modeling paradigm which mixes elements of continuous linear wave propagation and discrete signal processing in a way that allows knowledge of the analog properties of hyperbolic systems to reach powerful \textit{a priori} guarantees linking the temporal and spatial aspects of computations on graphs. This model is one of a family of biologically inspired computing approached including the Bat algorithm\cite{4} and Water Wave optimization\cite{5}, all of these approaches have narratively and conceptually compelling approaches to computational problems. We feel the SPIDER model's compelling combination of rigorous foundation with computationally exploitable properties, and extremely simple signal representation options may provide a computational framework which has different strengths and weaknesses than other approaches. We have already created basic implementations of this model for solving the hamiltonian cycle problem, and the traveling salesman problem, and continue to benchmark the performance of the algorithm on these problems.

\end{document}